# Erosion Study of Tungsten Carbide films under 100 keV Kr⁺ ion irradiation

Shristi Bist[1], Ratnesh K. Pandey[*,1], Sejal Shah[2,3], Parswajit Kalita[1], Amit Chawla[1], D. K. Avasthi[*1, 4]

[1]Department of Physics, School of Engineering, University of Petroleum and Energy Studies, Dehradun - 248007, Uttarakhand, India

[2]ITER India, Institute for Plasma Research, Bhat, Gandhinagar-382428, India

[3]Homi Bhabha National Institute (HBNI), Anushaktinagar, Mumbai-400094, India

[4]Centre for Interdisciplinary Research and Innovation, University of Petroleum and Energy Studies, Dehradun-248007, Uttarakhand, India

*Email: rpandey@ddn.upes.ac.in; dkavasthi@ddn.upes.ac.in

## Abstract

Tungsten carbide (WC) stands out as a crucial material for exploration in extreme environments due to its resistance to radiation and impressive mechanical strength. Widely utilized in cutting tools, high-wear components, and as a potential contender for plasma-facing material in nuclear reactors, WC's erosion behavior under surrogate irradiations is a subject of investigation. In the present work, WC films were synthesized at two different substrate temperatures of 400 K and 600 K using RF sputtering and were then irradiated with 100 keV Kr$^{1+}$ ions at a fluence of $1 \times 10^{17}$ ions/cm$^2$. The crystalline phases of as deposited WC films were confirmed by glancing incidence X-ray diffraction (GIXRD) measurements. Rutherford Backscattering Spectrometry (RBS) was employed to determine the thicknesses of pristine samples and the sputtering rate by measuring the difference in the areal densities of the pristine and irradiated films. The erosion rate of both types of films was found to be ~ 1.6 atoms per incident Kr⁺ ion. These findings contribute to a foundational comprehension of the radiation tolerance behavior of WC thin films, crucial for their performance in the demanding conditions of extreme radiation.

## Introduction

Tungsten Carbide (WC) is a well-known material for its exceptional hardness, wear resistance, and strength, making it a crucial material in various applications[1,2]. It has gained prominence in industries such as manufacturing, mining, oil and gas, and aerospace, owing to its exceptional mechanical properties[3–8]. Scientists are now studying its application as a plasma facing component (PFC)[9,10] in a nuclear reactor due to its extraordinary properties including high melting point, thermal conductivity and mechanical strength[7,11]. All these features make WC an attractive choice for withstanding the harsh environment of a plasma reactor. The first wall of a plasma reactor is subjected to radiation damage and sputtering due to the impact of energetic ions (hydrogen, helium, neutrons, etc.)[12–14]. Sputtering is a fundamental process in material science that occurs when energetic ions collide with a solid surface, causing the ejection of atoms and/or molecules from that surface[15,16]. The sputtering rate depends on ion energy[17,18] and also on various film parameters such as thickness[19], grain size[20], and substrate[21]. This process has significant implications in diverse fields including microelectronics, optics, material science and surface engineering [15,22–27]. In a recent paper by H. Tu et. al.[28], a study on erosion and retention properties of α-WC films by low energy deuterium ion irradiation was carried out. They pointed out that temperature played a little role in the W sputtering yield of WC and the erosion of WC by incident ions could be primarily described by physical sputtering. Additionally, preferential sputtering of C was also reported by them.

In the present work, the sputtering study of WC films deposited at two different substrate temperatures of 400 K and 600 K has been carried out under 100 KeV $Kr^+$ ion irradiation at a fluence of $1 \times 10^{17}$ ions/ $cm^2$. These investigations are essential for understanding the erosion dynamics of WC, particularly in the context of its applications in extreme environments. The results provide valuable insights into the material's response to radiation, crucial for optimizing its performance in cutting-edge applications such as nuclear reactors, where resistance to radiation and mechanical robustness are paramount.

**Materials and Methods**

Tungsten carbide films were meticulously deposited onto silicon (Si) substrates using an RF Magnetron sputtering setup[29] maintaining substrate temperatures of 400 K and 600 K. Before the deposition, the substrates were thoroughly cleaned using standard procedures in order to avoid contamination and to get a uniform film of high purity. A base vacuum of $4.9 \times 10^{-6}$ mbar was upheld in the sputtering chamber before initiating the deposition process. Sputtering was conducted under a pressure of $3 \times 10^{-2}$ mbar at room temperature, employing a 150 W RF power, and lasted for 30 minutes, with Argon gas flowing at a rate of 30 sccm.

To investigate the crystalline state of the pristine samples, glancing incidence X-ray diffraction (GIXRD) measurements were performed using advanced Empyrean X-Ray diffractometer (procured from Panalytical, Netherlands). The GIXRD pattern was recorded at

1° glancing angle using Cu K$_α$ radiation having wavelength equal to 1.54 Å, 45 kV generator voltage, and 40 mA tube current for the 2θ range from 20° to 80°.

Later, the as-deposited films were irradiated with 100 keV Kr$^+$ ions at normal incidence from ECR ion source based low energy ion beam facility at IUAC, New Delhi at fluence of 1 × 10$^{17}$ ions/cm$^2$ scanning half of the area (0.5cm × 1cm) of the samples. One half of the film was covered with Aluminium foil and the other half was exposed to ion irradiation as shown in fig. 1. Pristine as well as irradiated part of the films were examined through Rutherford Backscattering spectrometry (RBS) measurements using a 2 MeV He-ion beam from 1.7 MV tandem Pelletron (IUAC). The scattering angle during the RBS measurement was 165° and the beam spot size was ~ 1mm in diameter. The RBS spectrum has been analysed by using the SIMNRA 6.06 software[30].

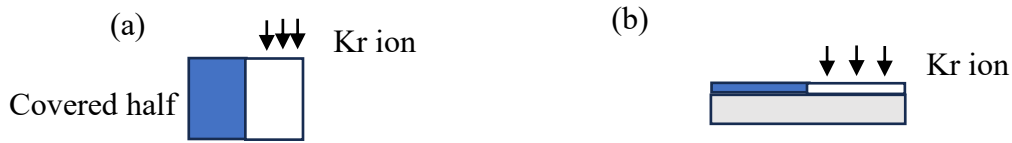

Fig.1 Schematic of the half-irradiated films (a) top view; (b) cross sectional view

**Results and Discussions**

GIXRD pattern of the pristine films is shown in fig. 2. The patterns reveal that the as-deposited films are polycrystalline in nature showing reflections at 2θ values of 36.81°, 62.05° and 74.4° corresponding to (1 0 0), (1 0 2), (1 1 0) and (2 0 0) planes of WC respectively (JCPDS-96-150-1543) and at 42.54° corresponding to (1 0 2) plane of W$_2$C (JCPDS-96-153-9793). The average grain size (D) of the films was calculated using Debye Scherrer's formula given by the equation: $D = \dfrac{k\lambda}{\beta \cos\theta}$, where 'k' is a constant (shape factor; k = 0.9), 'λ' is the wavelength of the X-rays, 'β' is full width at half maxima (FWHM) of the peak (in radians), and 'θ' is the angle corresponding to the peak (in degrees). The grain sizes were calculated to be ~5 nm and ~ 6 nm for films deposited at 400 K and 600 K substrate temperatures respectively.

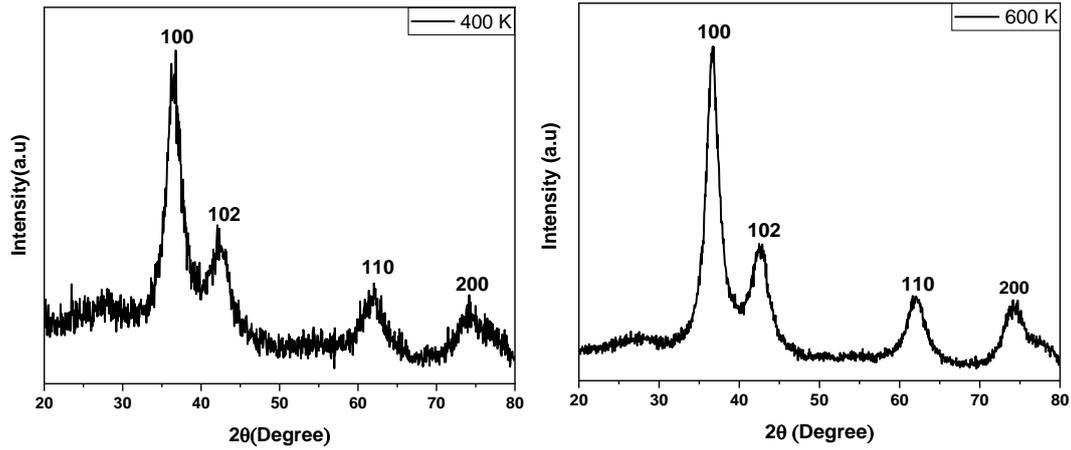

Fig. 2: GAXRD pattern of as-deposited WC films

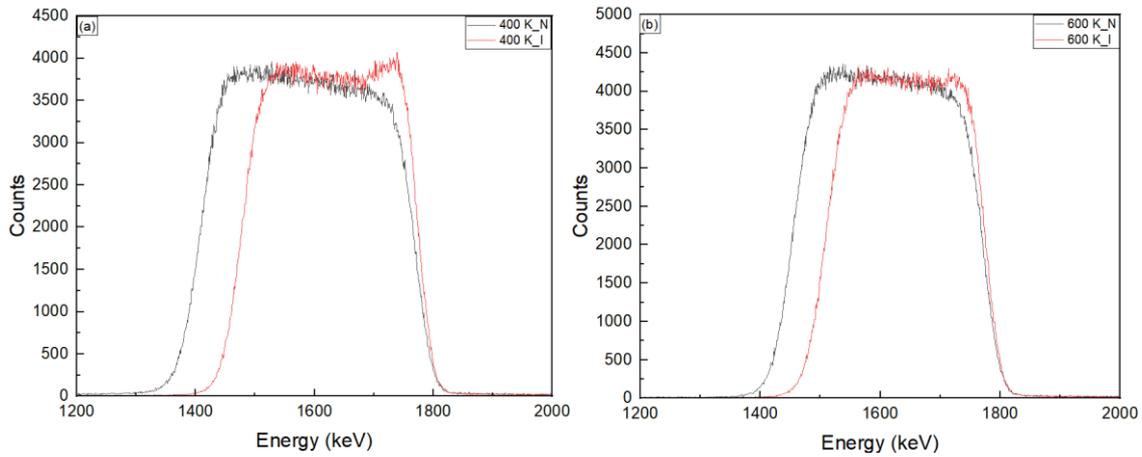

Fig. 3 spectra of W-peak of WC films deposited at (a) 400 K and (b) 600 K substrate temperature pristine (black) and irradiated with 100 keV Kr$^+$ ions at fluence $1 \times 10^{17}$ (red) ions/ cm$^2$

Figure 3 illustrates the tungsten (W) region of the RBS spectra for tungsten carbide (WC) films deposited at substrate temperatures of (a) 400 K and (b) 600 K. The analysis of the figures yield the following observations and conclusions:

(i) The reduction in the width of the W peak signifies the loss of tungsten due to ion irradiation.

(ii) A slight increase in the height of the W peak upon ion irradiation suggests the sputtering of lighter element C.

(iii) A distinctive rise in the peak height at energy values around 1800 keV indicates that, near the surface, there is an intensified sputtering of lighter elements, leading to a relative increase in tungsten concentration.

The second and third observations point towards the potential preferential sputtering of lighter elements induced by incident Kr ions.

It can also be seen that the extent of preferential sputtering in case of 400 K deposited film is much greater than that of 600 K deposited film, which may be because of the better stability of the W and C stoichiometry in WC film in case of film deposited at higher substrate temperature.

The theory of sputtering was first given out by Peter Sigmund[16]. According to it, the sputtering of target material by energetic incident ions could be due to the atomic collisions cascades. As a result of it erosion of the target material occur which can be characterized in the form of sputtering yield. Sputtering yield is typically the mean number of emitted atoms per incident particle.

Further, the sputter rate is calculated using the following relation:

$$\text{Sputter Rate} = \frac{Change\ in\ areal\ concentration}{Change\ in\ fluence} = \frac{\Delta c}{\Delta \phi}$$

Where,

Change in areal concentration = (Difference in the areal concentrations of non-irradiated and irradiated halves)

Fig. 4 and 5 show the experimental and simulated RBS spectra corresponding to the W peak of films deposited at 400 K and 600 K substrate temperatures. The sputtering measurements of the WC films were done by fitting the experimental data using SIMNRA 6.06 software. The calculated sputter rate for both type of samples is displayed in table 2. The grain size calculated using Debye Sherrer formula is also displayed in table 2

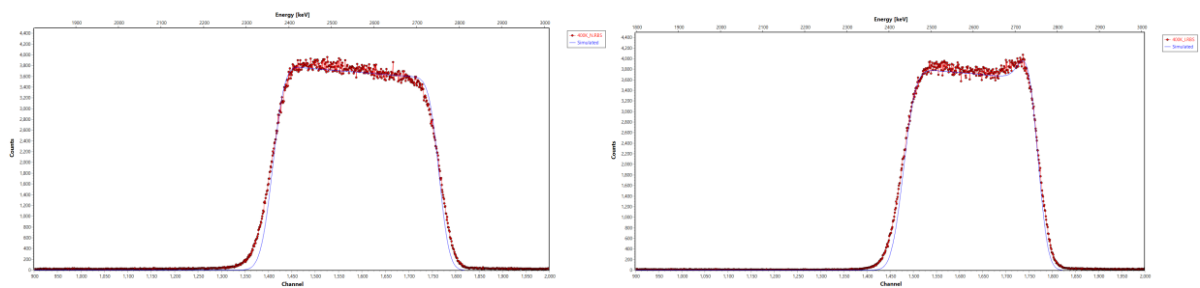

Fig. 4 Experimental RBS spectra (red) and the simulated spectra (blue), using SIMNRA 6.06, of as deposited (left) and irradiated (right) halves of 400 K WC film.

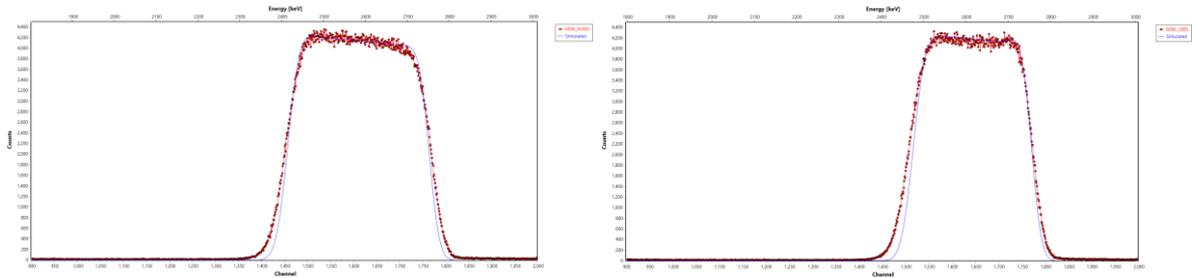

Fig. 5 Experimental RBS spectra (red) and the simulated spectra (blue), using SIMNRA 6.06, of as deposited (left) and irradiated (right) halves of 600 K WC film.

When energetic ion interacts with deposited films two processes occus: (i) elastic collisions leading to the fromation of collision cascades; and (ii) inelastic collisions leading to the formation of ion tracks. Elastic collisions are quantified by nuclear energy loss ($S_n$). In case of low energy ion irradiation, nuclear energy loss dominates due to the nature of interactions between the incident ion and target material which can be estimated using SRIM/ TRIM simulations in form of nuclear energy loss value ($S_n$). Stopping Range table was calculated using SRIM and it comes out to be < 50 nm for our case.
Theoretically, sputtering yield was also calculated while running TRIM simulation and comes out to be ~ 4.4 atom/ ion for W atom. However, from experimental RBS data, W sputter rate is calculated to be 1.6 atom/ ion. This value mismatched is due to the fact that TRIM overestimates the defects occurring due to ion irradiation.

Table 2.

| S. No. | Sample | Grain Size (nm) | Sputter Rate (atom/ ion) | |
|---|---|---|---|---|
| | | | Experimental | Theoretical |
| 1. | WC 400 K | ~ 5 | ~1.6 | ~4.4 |
| 2. | WC 600 K | ~ 6 | ~1.6 | ~4.4 |

However, we cannot comment anything regarding the preferential sputtering from TRIM simulation as only overall sputtering can be estimated by it. Preferential sputtering and dominant metallic phase can be observed from RBS experimental data (figures 3 and 4). Due to similar grain size of both type of the films (within the error bar) the sputter rate came out to be the same as the overall sputtering behaviour is significantly influenced by the grain size of the target material.[19,31]

## Conclusion

WC is being considered as a plasma-facing component in fusion reactor. It offers a unique combination of thermal resiliance and mechanical strength. In this paper, study of ersosion of WC films is carried out due to the effects of 100 keV $Kr^{1+}$ ion irradiation prepared via RF sputtering at substrate temperatures 400 K and 600 K. The results showed that as-deposited films are poly-crystalline and phases of WC and $W_2C$ are obtained as evident by GIXRD measurements. Preferential sputtering of C by Kr ion is observed leading to the appearance of dominant metallic phase on the WC films as evident from origin spectra.

This work provides a detailed study of sputtering of WC and will be helpful to predict the erosion behaviour in Tokamaks. With the advancement of plasma energy, WC plays an indespensable role as a key material in materials science in order to explore clean and sustainable energy sources.

## Acknowledgements

The authors, Shristi Bist, Ratnesh. K. Pandey and D. K. Avasthi are grateful to the Department of Atomic Energy, Board of Research in Nuclear Sciences (DAE-BRNS) for funding the research with sanction number 58/14/18/2021. Support from Inter University Accelerator Centre, New Delhi is highly acknowledged for beam time experiments and RBS characterizations. The support by UPES for RF/DC magnetron sputtering setup and thin film XRD at Central Instrumentation Centre (CIC) is gratefully acknowledged.

## References

1. García, J., Collado Ciprés, V., Blomqvist, A. & Kaplan, B. Cemented carbide microstructures: a review. *Int. J. Refract. Met. Hard Mater.* **80**, 40–68 (2019).

2. Badaluddin, N. A., Mohamed, I. F. & Ghani, J. A. Coatings of Cutting Tools and Their Contribution to Improve Mechanical Properties: A Brief Review. **13**, 12 (2018).

3. Humphry-Baker, S. A. & Smith, G. D. W. Shielding materials in the compact spherical tokamak. *Philos. Trans. R. Soc. Math. Phys. Eng. Sci.* **377**, 20170443 (2019).

4. Prakash, L. J. Application of fine grained tungsten carbide based cemented carbides. *Int. J. Refract. Met. Hard Mater.* **13**, 257–264 (1995).

5. Hussain, A., Podgursky, V., Antonov, M., Abbas, M. M. & Awan, M. R. Tungsten carbide material tribology and circular economy relationship in polymer and composites industries. *Proc. Inst. Mech. Eng. Part J. Mater. Des. Appl.* **236**, 2066–2073 (2022).

6. Korzhyk, V. N. *et al.* New Equipment for Production of Super Hard Spherical Tungsten Carbide and other High-Melting Compounds Using the Method of Plasma Atomization of Rotating Billet. *Mater. Sci. Forum* **898**, 1485–1497 (2017).

7. Liu, K. *Tungsten Carbide: Processing and Applications*. (BoD – Books on Demand, 2012).

8. Upadhyaya, G. S. *Cemented Tungsten Carbides: Production, Properties and Testing*. (William Andrew, 1998).

9. Doerner, R. P. The implications of mixed-material plasma-facing surfaces in ITER. *J. Nucl. Mater.* **363–365**, 32–40 (2007).

10. Garrison, L. M. *et al.* Review of Recent Progress in Plasma-Facing Material Joints and Composites in the FRONTIER U.S.-Japan Collaboration. *Fusion Sci. Technol.* **79**, 662–670 (2023).

11. Kurlov, A. S. & Gusev, A. I. *Tungsten Carbides: Structure, Properties and Application in Hardmetals*. vol. 184 (Springer International Publishing, 2013).

12. Jafari, A., Fayaz, V., Meshkani, S. & Terohid, S. A. A. Interaction Between Plasma and Tungsten Carbide Thin Films Coated on Stainless Steel as Tokamak Reactor First Wall. **147**, 7 (2018).

13. Jamal AbuAlRoos, N., Azman, M. N., Baharul Amin, N. A. & Zainon, R. Tungsten-based material as promising new lead-free gamma radiation shielding material in nuclear medicine. *Phys. Med.* **78**, 48–57 (2020).

14. Windsor, C. G. *et al.* Design of cemented tungsten carbide and boride-containing shields for a fusion power plant. *Nucl. Fusion* **58**, 076014 (2018).

15. Swann, S. Magnetron sputtering. *Phys. Technol.* **19**, 67–75 (1988).

16. Sigmund, P. Theory of Sputtering. I. Sputtering Yield of Amorphous and Polycrystalline Targets. *Phys. Rev.* **184**, 383–416 (1969).

17. Kumar, M. *et al.* Surface engineering of Pt thin films by low energy heavy ion irradiation. *Appl. Surf. Sci.* **540**, 148338 (2021).

18. Gupta, A. & Avasthi, D. K. Large electronically mediated sputtering in gold films. *Phys. Rev. B* **64**, 155407 (2001).

19. Pandey, R. K. *et al.* Study of electronic sputtering of CaF2 thin films. *Appl. Surf. Sci.* **289**, 77–80 (2014).

20. Pandey, R. K. *et al.* Swift heavy-ions induced sputtering in BaF2 thin films. *Nucl. Instrum. Methods Phys. Res. Sect. B Beam Interact. Mater. At.* **314**, 21–25 (2013).

21. Pandey, R. K. *et al.* Surface erosion of BaF2 thin films under SHI irradiation: Angular distribution and role of different substrates. *Appl. Surf. Sci.* **551**, 149343 (2021).

22. Bhatt, V. & Chandra, S. Silicon dioxide films by RF sputtering for microelectronic and MEMS applications. *J. Micromechanics Microengineering* **17**, 1066–1077 (2007).

23. Schiller, S., Heisig, U., Goedicke, K., Bilz, H. & Steinfelder, K. Methods and applications of plasmatron high rate sputtering in microelectronics, hybrid microelectronics and electronics. *Thin Solid Films* **92**, 81–98 (1982).

24. Edlou, S. M., Smajkiewicz, A. & Al-Jumaily, G. A. Optical properties and environmental stability of oxide coatings deposited by reactive sputtering. *Appl. Opt.* **32**, 5601 (1993).

25. Ferreira, A. A., Silva, F. J. G., Pinto, A. G. & Sousa, V. F. C. Characterization of Thin Chromium Coatings Produced by PVD Sputtering for Optical Applications. *Coatings* **11**, 215 (2021).

26. Murty, M. V. R. Sputtering: the material erosion tool. *Surf. Sci.* **500**, 523–544 (2002).

27. Musil, J. & Vlček, J. A perspective of magnetron sputtering in surface engineering. *Surf. Coat. Technol.* **112**, 162–169 (1999).

28. Tu, H., Li, C. & Shi, L. The erosion and retention properties of α-WC films by low-energy deuterium ion irradiation. *Appl. Surf. Sci.* **608**, 155133 (2023).

29. Bist, S. *et al.* Behavior of Tungsten Carbide thin Films Grown at Different Substrate Temperatures. in *2022 International Conference on Electrical Engineering and Photonics (EExPolytech)* 297–300 (2022). doi:10.1109/EExPolytech56308.2022.9951001.

30. Mayer, M. SIMNRA, a simulation program for the analysis of NRA, RBS and ERDA. *AIP Conf. Proc.* **475**, 541–544 (1999).

31. Kalita, P. *et al.* Radiation response of nano-crystalline cubic Zirconia: Comparison between nuclear energy loss and electronic energy loss regimes. *Nucl. Instrum. Methods Phys. Res. Sect. B Beam Interact. Mater. At.* **435**, 19–24 (2018).